\documentclass[aps,prb,amsmath,amssymb,superscriptaddress,preprint,floatfix,showpacs]{revtex4-2}

\usepackage{graphicx}
\usepackage[T3,T1]{fontenc}
\usepackage[cal=boondox,scr=boondoxo]{mathalfa}
\usepackage{subfigure}
\usepackage{mathtools}
\usepackage[mathscr]{euscript}
\usepackage{slashed}
\usepackage{hyperref}
\usepackage{color}

\newcommand{\beq}{\begin{equation}}
\newcommand{\eeq}{\end{equation}}
\newcommand{\rf}[1]{(\ref{#1})}
\newcommand{\sdot}{\!\cdot\!}

\def\etal{{\it et al.}}

\makeatother

\begin{document}

	\title{Nonperturbative Lorentz Violation and Field Quantization}
	
	\author{V.\ Alan Kosteleck\'y}
	\affiliation{Department of Physics,
		Indiana University,
		Bloomington,
		Indiana 47405,
		USA}
	\affiliation{Indiana University Center for Spacetime Symmetries,
		Bloomington, Indiana 47405, USA}
	
	\author{Ralf Lehnert}
	\affiliation{Department of Physics,
		Indiana University,
		Bloomington,
		Indiana 47405,
		USA}
	\affiliation{Indiana University Center for Spacetime Symmetries,
		Bloomington, Indiana 47405, USA}
		
	\author{Marco Schreck}
	\affiliation{Departamento de F\'{i}sica, 
		Universidade Federal do Maranh\~{a}o,
		Campus Universit\'{a}rio do Bacanga,
		S\~{a}o Lu\'{i}s (MA),
		65085-580,
		Brazil}
	
	\author{Babak Seradjeh}
	\affiliation{Department of Physics,
		Indiana University,
		Bloomington,
		Indiana 47405,
		USA}
	\affiliation{Indiana University Center for Spacetime Symmetries,
		Bloomington, Indiana 47405, USA}
	\affiliation{Quantum Science and Engineering Center,
		Indiana University,
		Bloomington,
		Indiana 47405,
		USA}
		
	\date{December 2024}
		
		\begin{abstract}
			Regimes of Lorentz-violating effective field theories 
			are studied
			in which departures from Lorentz symmetry 
			are nonperturbative.
			Within a free toy theory 
			exhibiting Lorentz breakdown involving an operator of mass dimension three,
			it is shown 
			that conventional methods suffice 
			to achieve field quantization 
			and Fock-space construction.
			However, 
			the absence of an observer-invariant 
			energy-positivity condition 
			requires physical input beyond the free theory 
			for the unambiguous identification of a ground state. 
			An investigation of the role of thermodynamics 
			in this context is instigated.
		\end{abstract} 
		
		
		{
			\let\clearpage\relax
			\maketitle
		}
		
		\newpage
	
	\section{Introduction}
	
	With its outsize role in both 
	the Standard Model and General Relativity, 
	Lorentz invariance has long been identified as a centerpiece 
	of established physics.
	Probing the validity of this foundational symmetry 
	therefore represents a guidepost 
	for robust progress 
	in fundamental-physics research. 
	At the same time,
	departures from this symmetry 
	can be accommodated 
	in various approaches to new physics~\cite{ks89,kp91}
	providing an additional premise 
	for tests of Lorentz invariance.
	
	Phenomenological Lorentz-violation studies 
	are facilitated 
	by a framework based on effective field theory~\cite{sw} 
	containing the Standard Model and General Relativity,
	known as the Standard-Model Extension (SME)~\cite{kp95,ck97,ck98,ak04,kl21}.
	The SME 
	classifies 
	the various types of Lorentz-breaking effects
	and allows the identification,
	interpretation, and comparison 
	of feasible Lorentz tests. 
	To date,
	this framework has formed the basis
	for numerous experimental studies 
	of Lorentz symmetry~\cite{tables}.
	
	In most applications of the SME,
	the Lorentz violation is presumed perturbatively small, 
	a property largely dictated by observational evidence.
	While this feature 
	simplifies phenomenological predictions,
	it may at times obscure conceptual insight. 
	Consider energy positivity,
	for example,
	which is a property commonly invoked 
	to evaluate the viability of quantum field theories (QFTs). 
	In the SME,
	this property holds 
	for sufficiently small Lorentz-violating coefficients. 
	However,
	these coefficients are components of Lorentz tensors 
	and may thus be large 
	for highly boosted observers. 
	For such observers,
	Lorentz violation appears nonperturbative,
	and energy positivity typically fails~\cite{kl01}.
	This behavior 
	seems to indicate the undesirable prediction 
	that observers in different states of motion 
	are unable to agree 
	on physical implications of the SME.
	The apparent paradox
	represents a key aspect 
	of the longstanding concordance problem~\cite{kl01,WeylConcordance}.
	
	This work obtains conceptual insight
	into these questions
	via an analysis   
	of the Lorentz-violating $b_\mu$ theory~\cite{ck97,ck98},
	which enjoys widespread popularity in physics.
	With perturbatively small background-field coefficients $b_\mu$,
	it has been the basis for numerous phenomenological SME studies
	involving electrons, protons, neutrons, and other fermions~\cite{bclms24,p1,p2,p3,p4,p5,p6,p7,p8,p9,ms17,p10,p16,bkl16,p11,p12,p13,p14,p15,p17,p18,p19,p20,p21,p22,p23,p24,p25,p26,p27,p28,p29,p30,p31,p32,p33,p34,p35,p36,p37} 
	and has provided a theoretical laboratory 
	for investigations of Lorentz violation in fundamental QFT~\cite{ck97,ck98,kl01,jk99,mp99,co99,jc99,jc992,klp02,ba04,ba042,rl06,bt1,bt2,bt3,km13,bt4,bt5,rs17,bt6,bt7,bt9,pv23,bt10,klss25}. 
	The geometric underpinnings of the theory 
	can be characterized 
	by Finsler geometry~\cite{ak11,f1,js13,f2,f3,cm15,f4,f5,f6}.
	In condensed-matter physics, 
	the $b_\mu$ theory
	governs the band structures of certain semimetals 
	near Weyl nodes~\cite{ag12,zwb12,s1,kk16,rj16,gc16,kw17,jw17,nx17,bskrg19,lqd21,s2,s3,klmss22,s4,gdmu24}.
	
	Our study
	takes the Lorentz-violating $b_\mu$ to be nonperturbatively large 
	for all observers,
	leading to negative energies 
	in essentially all frames. 
	This approach 
	disentangles issues arising 
	due to negative energies
	from those due to coordinate changes.
	Owing to the linearity of the theory,
	the single-particle quantum-mechanics version 
	is exactly solvable.
	We expose the mild assumptions needed 
	to perform second quantization
	and Fock-space construction,
	and we demonstrate 
	that the usual criteria 
	for ground-state selection
	become ambiguous.
	Further physical input 
	beyond the free theory 
	is required, 
	and we show 
	that thermodynamical reasoning
	can play a key role in this respect.
	
	This paper is organized as follows.
	In Sec.~\ref{basics},
	we introduce our toy theory 
	and recall some of its basic properties.
	Section~\ref{branches} 
	characterizes the plane-wave dispersion-relation branches
	of this theory
	before second quantization.
	Canonical quantization of the model 
	based on a Dirac constraint analysis 
	is performed 
	and the corresponding Fock space is constructed 
	in Sec.~\ref{quant}.
	Section~\ref{vacuum} 
	comments on the identification 
	of a suitable ground state 
	in the Fock space of the theory.
	Throughout,
	we work in natural units 
	$\hbar=c=k_B=1$, 
	and the Minkowski metric $\eta_{\mu\nu}$ 
	has signature $(+,-,-,-)$.

	\section{Model basics}
	\label{basics}
	
	The Lagrange density of the free $b_{\mu}$ theory is~\cite{ck97,ck98}
	\begin{equation}
		\label{eq:Dirac-theory-modified-b}
		\mathcal{L}_b=\tfrac{1}{2}\bar{\psi}\big(\mathrm{i}\mkern+1mu\slashed{\phantom{b}}\mkern-9.5mu\partial-m-\gamma^5\slashed{\phantom{b}}\mkern-10mu b\big)\psi+\text{h.c.}\,,
	\end{equation}
	where $\psi$ is a spin-$\tfrac{1}{2}$ fermion.
	We follow the gamma-matrix conventions  
	of Ref.~\cite{iz80}.
	The nondynamical real-valued $b_{\mu}$
	selects a preferred spacetime direction 
	violating both Lorentz and CPT invariance. 
	We will take $b_{\mu}$ as spacelike,
	a choice that is both convenient 
	and physically realized in certain Weyl semimetals~\cite{ag12,bskrg19}.
	
	In the present flat-spacetime context, 
	we take $b_{\mu}$ as constant in cartesian coordinates, 
	so that the canonical energy--momentum tensor $\theta^{\mkern+2mu\mu\nu}$
	is conserved 
	and permits the construction of a conserved 4-momentum $P^\nu$:
	\beq\label{EMT}
	\theta^{\mkern+2mu\mu\nu}=\tfrac{\mathrm{i}}{2}\bar{\psi}\mkern+2mu\gamma^{\mkern+2mu\mu}\partial^{\nu}\psi 
	+\textrm{h.c.}\,,\qquad P^{\nu}=\int \! \mathrm{d}^3\mkern-2mu x\;\theta^{\mkern+2mu 0\nu}\,.
	\eeq
	The theory~\rf{eq:Dirac-theory-modified-b} also has a global U(1) symmetry 
	with a conserved current $j^{\mkern+2mu\mu}$ 
	giving a conserved charge $N$:
	\beq\label{consJ}
	j^{\mkern+2mu\mu}=\bar{\psi}\mkern+2mu\gamma^{\mkern+2mu\mu}\psi\,,\qquad N=\int \! \mathrm{d}^3\mkern-2mu x\;j^{\mkern+2mu 0}\,.
	\eeq
	In the QFT context to be discussed below, 
	$N$ may be interpreted as the fermion number, 
	as usual.
	
	The theory~\rf{eq:Dirac-theory-modified-b} yields the modified Dirac equation
	\beq\label{modDirac}
	\big(\mathrm{i}\mkern+2mu\slashed{\phantom{b}}\mkern-9.5mu\partial-m-\gamma^5\slashed{\phantom{b}}\mkern-10mu b\big)\psi=0\,.
	\eeq
	This equation can be cast into the Schr\"odinger form
	\beq\label{modDiracH}
	\mathrm{i}\partial^0\psi=H\psi\,,\qquad H\coloneqq\gamma^0(\mathrm{i}\mkern+2mu\vec{\gamma}\sdot\vec{\nabla}+m+\gamma^5\slashed{\phantom{b}}\mkern-10mu b)\,,
	\eeq
	where the hamiltonian $H$ is hermitian. 
	A plane-wave \textit{ansatz} $\psi(x)=\textrm{\sffamily\itshape w\hspace{0.3mm}}(\lambda)\exp(-\mathrm{i}\lambda\cdot x)$, 
	where $\lambda^{\mu}=(\omega,\vec{\lambda})$ is the wave 4-vector 
	and $\textrm{\sffamily\itshape w\hspace{0.3mm}}(\lambda)$ a 4-component spinor, 
	yields 
	\beq\label{DR}
	(\lambda^2-m^2-b^2)^2+4b^2\lambda^2-4(\lambda\cdot b)^2=0\,.
	\eeq
	This dispersion relation is a fourth-order polynomial 
	in the wave frequency $\omega$, 
	generating four solutions $\omega(\vec{\lambda})$. 
	Due to the hermiticity of $H$, 
	these four branches are real valued
	for any real $\vec{\lambda}$ 
	and any size of~$b_\mu$.

	\section{Dispersion-relation branch structure}
	\label{branches}
	
	Upon second quantization of the $b_\mu$ theory,
	part of the spectrum determined by Eq.~\rf{DR} 
	must be reinterpreted, 
	so $\lambda^\mu$ may not always represent the physical momentum. 
	This reinterpretation,
	which is closely tied to the separation of the $\lambda^\mu$ spectrum into branches,
	thus affects energy-positivity analyses.
	This section provides such a separation
	without giving preference to any coordinate system,
	i.e., based on coordinate-independent features.
	These features include topological connectedness 
	and smoothness of the group velocity $\vec{v}_{\rm g}$.
	
	We begin by characterizing the four branches 
	by their behavior as $|\vec{\lambda}|\to\infty$. 
	Direct calculation shows 
	that this asymptotic behavior 
	corresponds to the ultrarelativistic limit $m\to\infty$ 
	and is therefore governed by
	the condition $(\lambda+b)^2(\lambda-b)^2=0$.
	Asymptotically, 
	there are thus two positive- and two negative-valued branches, 
	as expected. 
	We denote these by $\omega^{\pm}_s(\vec{\lambda})$, 
	where the superscript $\pm$ labels the asymptotic behavior
	and $s$ the remaining twofold multiplicity 
	associated with the spin degree of freedom. 
	Our notation for the corresponding wave vectors and eigenspinors 
	is $(\lambda^{\pm}_s)^\mu=(\omega^{\pm}_s,\vec{\lambda})$ 
	and $\textrm{\sffamily\itshape w}^{\,\pm}_s(\vec{\lambda})$, 
	respectively.
	
	\begin{figure*}
		\subfigure[]{\label{fig:sfig1}\includegraphics[scale=0.22]{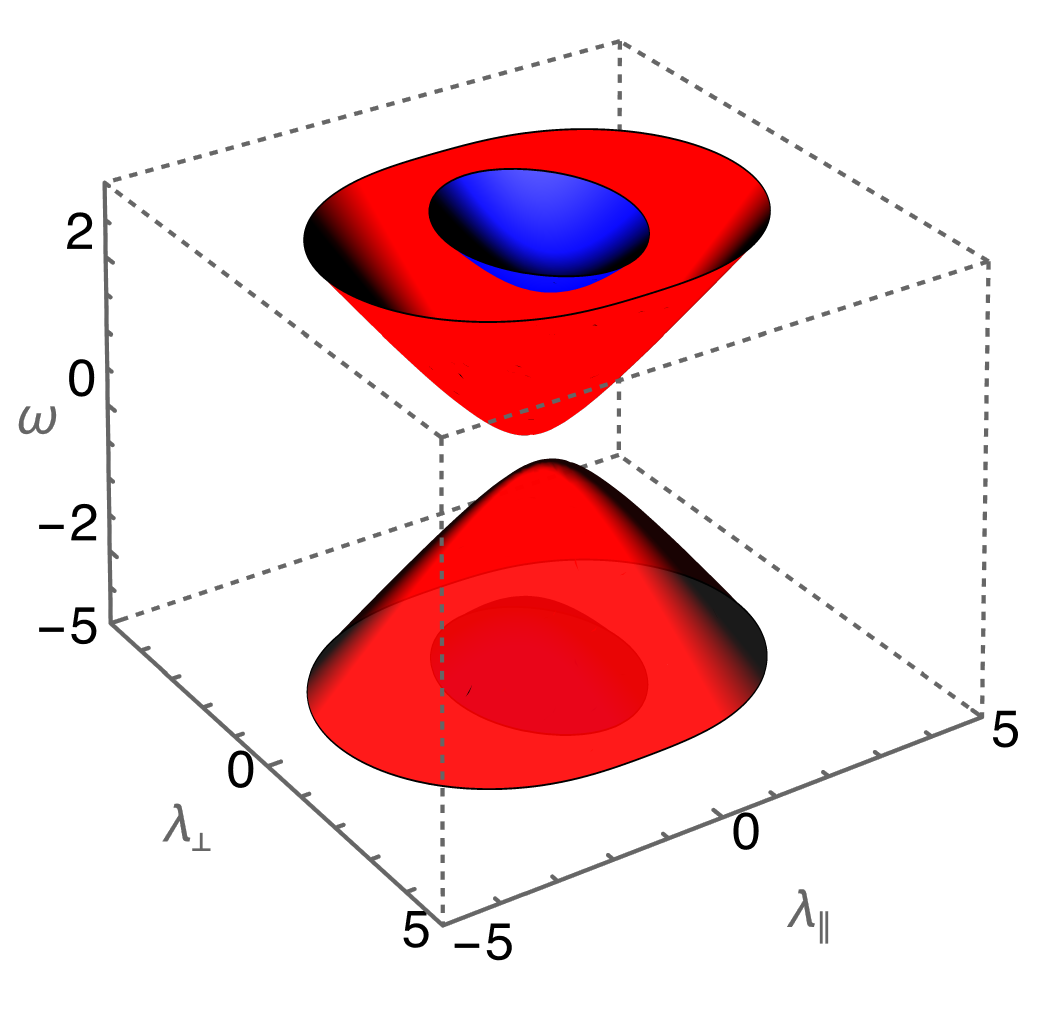}}
		\subfigure[]{\label{fig:sfig2}\includegraphics[scale=0.22]{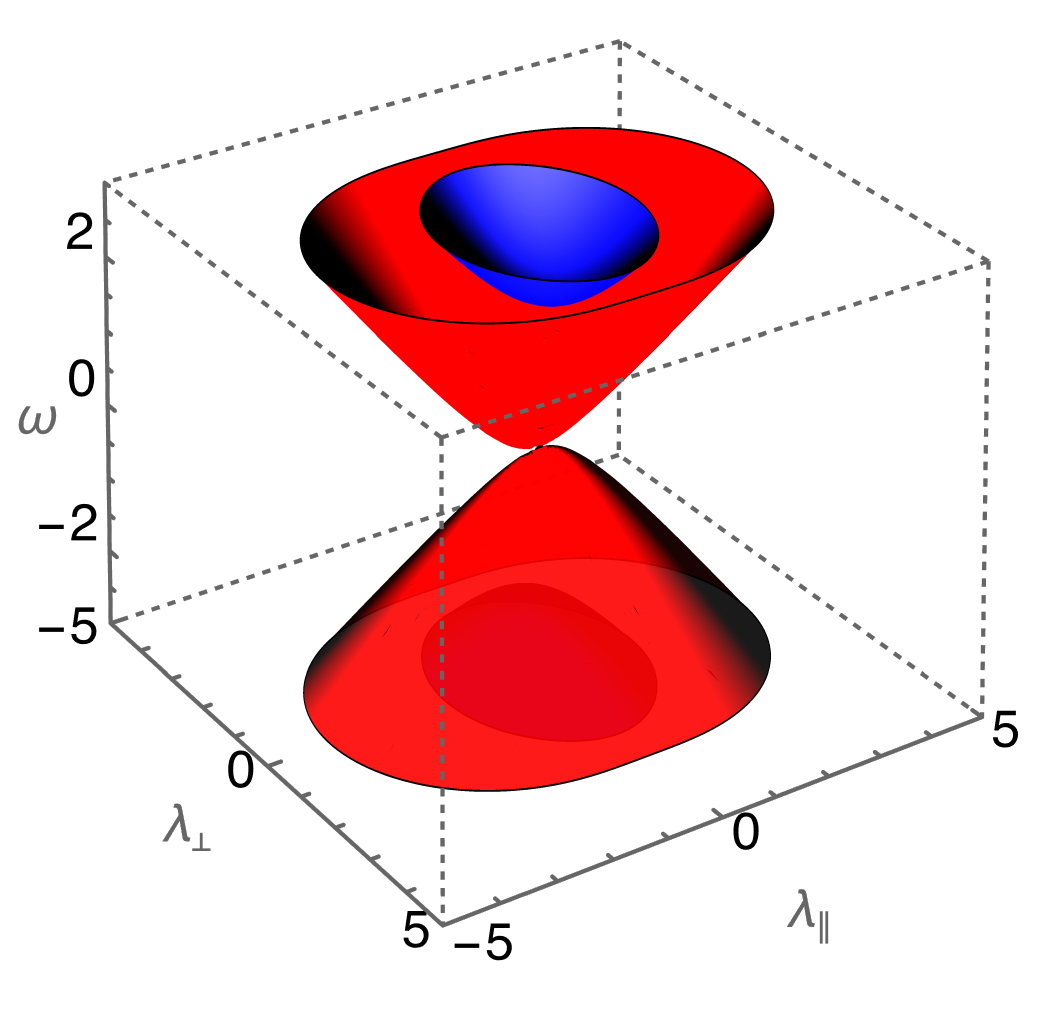}}
		\subfigure[]{\label{fig:sfig3}\includegraphics[scale=0.22]{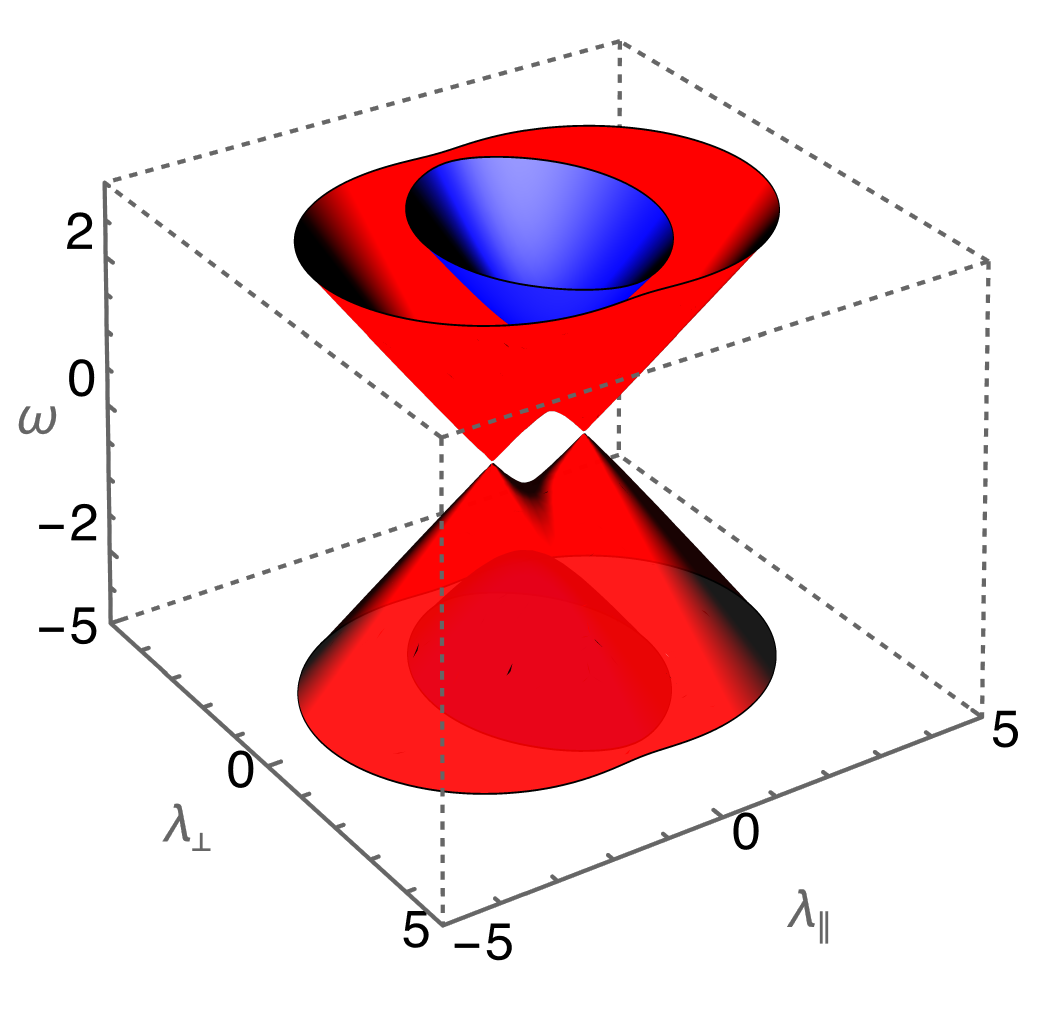}}
		\subfigure[]{\label{fig:sfig4}\includegraphics[scale=0.22]{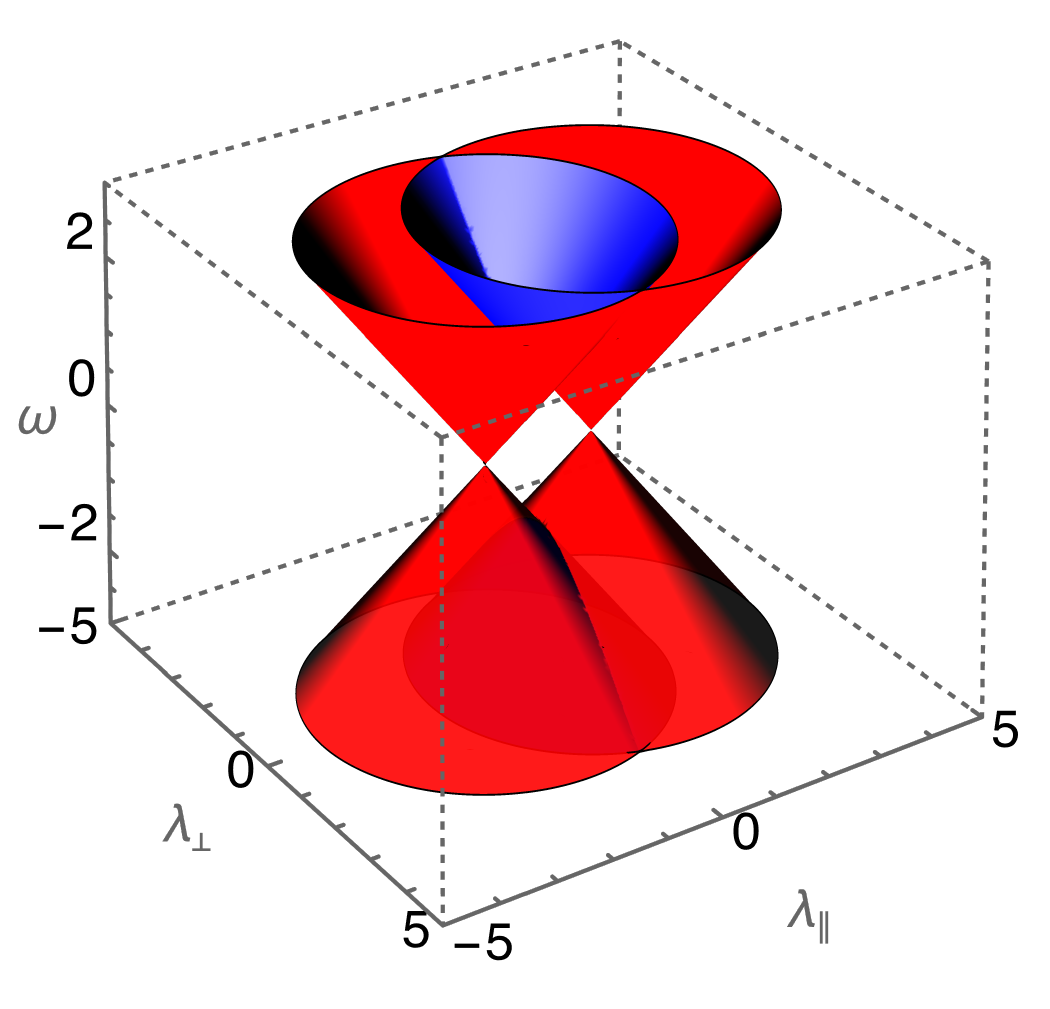}}
		\caption{Dispersion-relation branches for purely spacelike $b^\mu=(0,\vec{b})$. 
			The plane-wave frequencies $\omega^\pm_s(\vec{\lambda})$ 
			are plotted versus $\lambda_\parallel$ and $\lambda_\perp$,
			the components of $\vec{\lambda}$ parallel and perpendicular to $\vec{b}$. 
			The color red signifies $s=1$, 
			and blue corresponds to $s=2$.
			For large masses $m^2>-b^2$, 
			all four branches are separated (a); 
			for $m^2=-b^2$ the two $s=1$ branches touch at the single point 
			located at the origin of $\lambda^\mu$ space (b); 
			for small masses $m^2<-b^2$, 
			two isolated points of contact between the $s=1$ branches exist (c);
			in the $m\to 0$ limit, 
			two isolated points of contact between the $s=1$ branches persist,
			but the $s=1$ and $s=2$ branches are no longer separated (d).
		}
		\label{Fig1}
	\end{figure*}
	
	For the massive case $m\neq 0$,
	we select a label $s$ 
	based on the spin-type operator
	\beq\label{chi}
	\mathscr{S}\coloneqq-2\gamma^5\big(b\cdot \mathrm{i}\partial-m \mkern+2mu \slashed{\phantom{b}}\mkern-10mu b\big)\,,
	\eeq
	which obeys
	\beq\label{chi_comm}
	[\mathscr{S},H]=2m[\gamma^0,\gamma^5\slashed{\phantom{b}}\mkern-10mu b]
	\big(\mathrm{i}\mkern+1mu\slashed{\phantom{b}}\mkern-9.5mu\partial-m-\gamma^5\slashed{\phantom{b}}\mkern-10mu b\big)\,.
	\eeq
	On the space of solutions to our modified Dirac equation~(\ref{modDirac}), 
	$\mathscr{S}$ therefore commutes with the hamiltonian, 
	so that the plane-wave solutions $\textrm{\sffamily\itshape w}^{\,\pm}_s(\vec{\lambda})\exp (-\mathrm{i}\lambda^{\pm}_s\cdot x)$ 
	are also eigenvectors of $\mathscr{S}$. 
	The label $s$ may now be fixed 
	as parametrizing the sign of the $\mathscr{S}$ eigenvalues given by
	\beq\label{chi_eigen}
	S_{\mkern-4mu s}(\lambda^{\pm}_s)=2(-1)^s\,\sqrt{(b\cdot\lambda^{\pm}_s)^2-b^2m^2}\,,\quad s=1,2\,.
	\eeq
	To show 
	that $s$ indeed distinguishes topologically between same-sign roots of Eq.~\rf{DR}
	we express $S_{\mkern-4mu s}(\lambda^{\pm}_s)$ as 
	\beq\label{chi_eigen2}
	S_{\mkern-4mu s}(\lambda^{\pm}_s)=(\lambda^{\pm}_s)^2-m^2+b^2\;.
	\eeq
	Note that for spacelike $b_\mu$ and $m\neq 0$, 
	the square root in Eq.~\rf{chi_eigen} is positive definite. 
	Hence, 
	the hyperboloid $\lambda^2-m^2+b^2=0$ 
	strictly separates the $s=1$ and $s=2$ branches, 
	which have no intersection.
	
	To study topological relations 
	between opposite-sign roots, 
	we use continuity and bijectivity of the Lorentz transformations 
	to select a frame in which $b^\mu=(0,\vec{b})$ is purely spacelike. 
	Equation~(\ref{chi_eigen2}) then yields 
	\beq\label{spacelike_roots}
	\omega^{\pm}_s(\vec{\lambda})=\pm\sqrt{m^2+\vec{\lambda}^2+\vec{b}^2+S_{\mkern-4mu s}(\vec{\lambda})}\;.
	\eeq
	The structure of this expression shows that a plus and a minus branch 
	are always separated 
	unless the right-hand side of this equation vanishes. 
	This can occur only for the $s=1$ branches and requires
	\beq\label{spacelike_touching}
	\lambda_\mu=\pm\sqrt{1+\frac{m^2}{b^2}}\,b_\mu\,,
	\eeq
	where we have boosted back to the original frame, 
	and the $\pm$ signs are uncorrelated with those in Eq.~(\ref{spacelike_roots}). 
	The resulting topologies are depicted in Fig.~\ref{Fig1}.
	\begin{figure}
		\centering
		\includegraphics[width=0.3\linewidth]{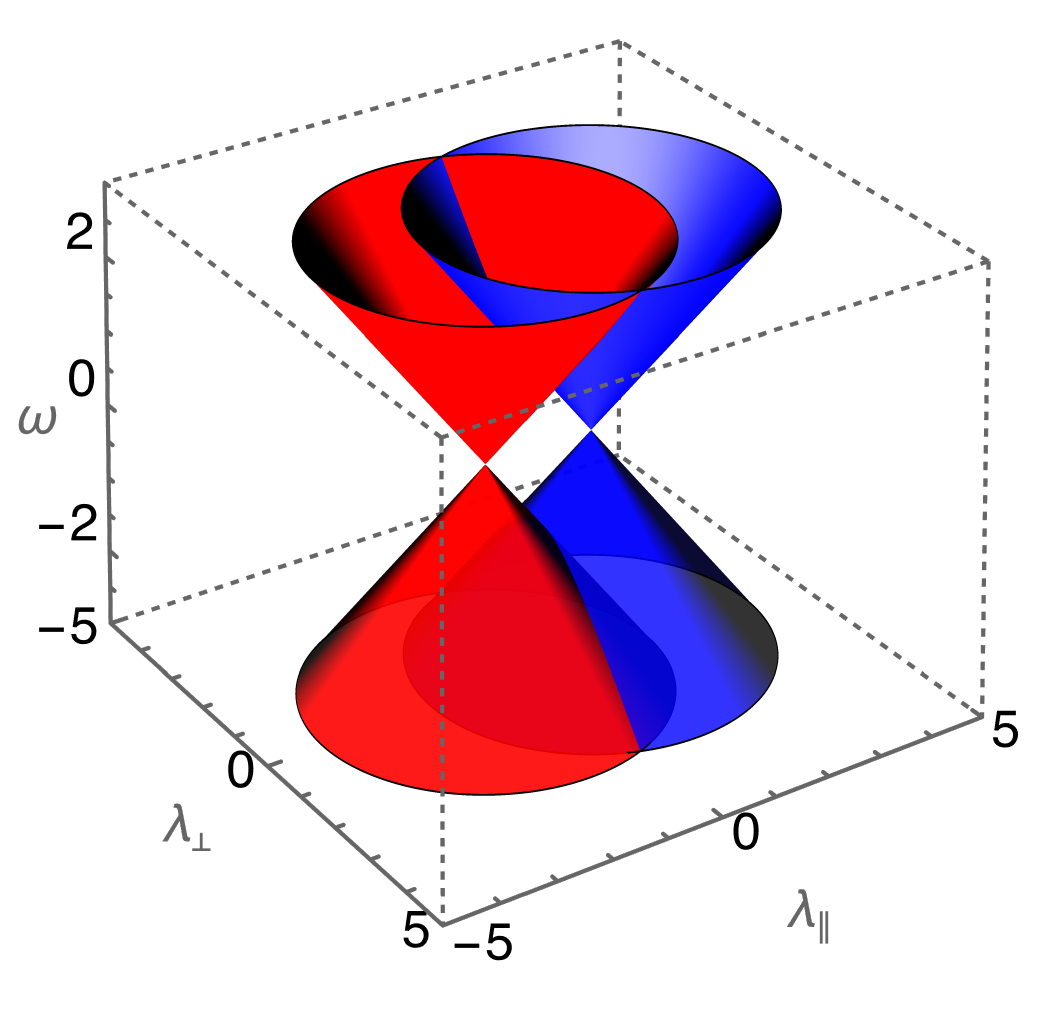}
		\caption{Dispersion-relation branches 
			for the massless case $m=0$. 
			This plot is based on the same $b_\mu$ value 
			as that in Fig.~\ref{fig:sfig4}.
			However, the branches are colored 
			according to their chirality,
			with $\chi=+1$ in red and $\chi=-1$ in blue.
		}
		\label{fig:chi}
	\end{figure}
	
	In the massless case $m=0$,
	the topology changes: 
	the $s=1$ and $s=2$ branches are no longer separated.
	To see this,
	note that the on-shell eigenvalues of $\mathscr{S}$ vanish 
	for plane-wave momenta satisfying $b\cdot\lambda^{\pm}_s=0$.
	The existence of such momenta may be established, 
	without loss of generality, 
	in a frame in which $b^\mu=(0,\vec{b})$ is purely spacelike: 
	$b\cdot\lambda^{\pm}_s=-\vec{b}\cdot\vec{\lambda}$ vanishes 
	for all $\vec{\lambda}$ perpendicular to $\vec{b}$. 
	At such momenta, 
	the $s=1$ and $s=2$ branches intersect, 
	as shown in Fig.~\ref{fig:sfig4}. 
	
	Figure~\ref{fig:sfig4} also indicates
	that $\vec{v}_{\rm g}$ on a given $s$ branch
	fails to change smoothly 
	across these branch intersections,
	as is verified below.
	For $m=0$, 
	we therefore select an alternative branch identification 
	based on the the ordinary chirality operator $\gamma^5$. 
	In the massless case, 
	this operator is indeed conserved 
	and possesses eigenvalues $\chi=\pm 1$.
	On plane-wave eigenspinors $\textrm{\sffamily\itshape w}^{\,\pm}_\chi(\vec{\lambda})$, 
	the $\gamma^5$ eigenvalues $\chi$ may also be expressed as
	\beq\label{chirality_eigen}
	\gamma^5\textrm{\sffamily\itshape w}^{\,\pm}_\chi(\vec{\lambda})=
	\chi
	\textrm{\sffamily\itshape w}^{\,\pm}_\chi(\vec{\lambda})
	=
	-\frac{(\lambda^\pm_\chi )^2+ b^2}{2\lambda^\pm_\chi \cdot b}\textrm{\sffamily\itshape w}^{\,\pm}_\chi(\vec{\lambda})\,,
	\eeq
	where it is understood 
	that this simply represents a relabeling 
	of our previous solutions 
	in terms of their chirality eigenvalue $\chi$
	instead of their $\mathscr{S}$ eigenvalue $s$. 
	The corresponding branch identification 
	is depicted in Fig.~\ref{fig:chi}.
	
	We proceed 
	with a discussion of the wave-packet group velocity 
	$\vec{v}_{\rm g}=\vec{\nabla}_{\mkern-4mu \vec{\lambda}}\mkern+3mu\omega(\vec{\lambda})$
	and its smoothness.
	For purely spacelike $b_{\mu}$, 
	the group velocity is
	\beq\label{groupv}
	\vec{v}^{\,\pm}_s(\vec{\lambda})=\pm\frac{1}{\omega^\pm_s(\vec{\lambda})}\left(
	\vec{\lambda}+2\frac{\vec{b}\cdot\vec{\lambda}}{S_{\mkern-4mu s}(\vec{\lambda})}\vec{b}
	\right)\,.
	\eeq
	For $m^2\leq -b^2$, 
	this expression becomes undefined 
	at the isolated touching points 
	determined by Eq.~(\ref{spacelike_touching})
	and apparent in Figs.~\ref{fig:sfig2}, \ref{fig:sfig3}, and \ref{fig:sfig4}.
	Paralleling the ordinary case of a lightlike dispersion relation, 
	we may disregard this indeterminacy 
	for our present analysis. 
	
	In the massive case $m\neq 0$,
	the group velocity~\rf{groupv}
	is otherwise well defined, continuous, and subluminal,
	\beq\label{group_v_bound}
	|\vec{v}^{\,\pm}_s(\vec{\lambda})|\leq 1\,.
	\eeq
	Employing again the continuity and bijectivity of the Lorentz transformations
	then establishes branch-wise smoothness
	for any spacelike $b_\mu$ in the massive case.
	
	In the massless case $m=0$,
	$S_{\mkern-4mu s}(\vec{\lambda})\to 2(-1)^s |\vec{b}\cdot\vec{\lambda}|$,
	so that the group velocity~\rf{groupv} 
	is discontinuous on $s$ branches 
	when $\vec{b}\cdot\vec{\lambda}$ 
	changes sign, 
	as noted above; see Fig.~\ref{fig:sfig4}.
	A chirality branch,
	on the other hand, 
	obeys $(\lambda+\chi b)^2=0$,
	as can be established using Eq.~(\ref{chirality_eigen}).
	The corresponding group velocities 
	\beq\label{group_v_massless_chi}
	\vec{v}^{\,\pm}_\chi(\vec{\lambda})=\pm\frac{\vec{\lambda}+\chi\vec{b}}{|\vec{\lambda}+\chi\vec{b}|}
	\eeq
	are now well defined; see Fig.~\ref{fig:chi}.
	Although these group velocities are not aligned with the wave 3-momenta, 
	superluminality is avoided because
	$|\vec{v}^{\,\pm}_\chi(\vec{\lambda})|=1$.
	Again, 
	these results can be extended to any spacelike $b_\mu$ 
	by the continuity and bijectivity properties of the Lorentz transformations.
	
	We conclude this section 
	with a few notes on eigenspinors 
	associated with the four branches.
	They obey
	\beq\label{spinor_norm}
	\textrm{\sffamily\itshape w}^{\,\varsigma\dagger}_r(\vec{\lambda})\,
	\textrm{\sffamily\itshape w}^{\,\varsigma'}_{r'}(\vec{\lambda})
	=N^\varsigma_r(\vec{\lambda})\,\delta_{rr'} \delta^{\varsigma\varsigma'}\,,
	\eeq
	where both spin-type quantum numbers $r$ and $r'$ 
	are either $\mathscr{S}$ labels $s$ if $m\neq 0$ 
	or chirality labels $\chi$ if $m=0$, 
	and $\varsigma,\varsigma'\in\{+,-\}$.
	Orthogonality~\rf{spinor_norm} 
	follows because the $\textrm{\sffamily\itshape w}^{\,\varsigma}_{r}(\vec{\lambda})$
	are eigenvectors of the hermitian hamiltonian~(\ref{modDiracH}) 
	corresponding to nondegenerate eigenvalue branches.
	At the branch intersections for $m=0$, 
	the $\textrm{\sffamily\itshape w}^{\,\varsigma}_{\chi}(\vec{\lambda})$ 
	remain orthogonal 
	because they are simultaneously nondegenerate chirality eigenvectors.
	
	Our choice of spinor normalization is
	\begin{equation}\label{spinor_normtwo}
		N^\pm_r(\vec{\lambda}) =
		\begin{cases}
			2m\sqrt{1-\vec{v}^{\,\pm}_s(\vec{\lambda})^2} & \text{for $m\neq 0$, $r\to s$}\\
			2|\vec{\lambda}+\chi \vec{b}| & \text{for $m=0$, $r\to \chi$}\,.\\
		\end{cases}       
	\end{equation}
	For $b_{\mu}\to 0$, 
	our normalizations agree 
	with those often chosen in the Lorentz-symmetric case.
	Note  
	that together with Eq.~\rf{modDirac} 
	the normalization~\rf{spinor_normtwo} implies
	\beq\label{vgr_spinor}
	N^\pm_r(\vec{\lambda})\vec{v}^{\,\pm}_r(\vec{\lambda})=
	\overline{\hspace{-1pt}\textrm{\sffamily\itshape w\hspace{+1pt}}}^{\,\pm}_r(\vec{\lambda})\,\vec{\gamma}\,
	\textrm{\sffamily\itshape w}^{\,\pm}_{r}(\vec{\lambda})\,
	\eeq
	i.e.,
	$\overline{\hspace{-1pt}\textrm{\sffamily\itshape w\hspace{+1pt}}}^{\,\pm}_r(\vec{\lambda})\,\gamma^{\mu}\,\textrm{\sffamily\itshape w}^{\,\pm}_{r}(\vec{\lambda})$ 
	transforms as a 4-vector 
	and $\textrm{\sffamily\itshape w}^{\,\pm}_{r}(\vec{\lambda})$ as a spinor 
	under coordinate changes.
	
	The general solution $\psi(x)$ to Eq.~(\ref{modDirac}) 
	can now be characterized as a superposition of plane waves:
	\begin{equation}\label{planewavedecomp}
		\psi(x)= \int \!\frac{\mathrm{d}^3\lambda}{(2\pi)^3}\sum_{r}
		\left[ 
		\frac{b_r(\vec{\lambda})}{N^+_r(\vec{\lambda})}
		\textrm{\sffamily\itshape w}^{+}_r(\vec{\lambda})
		\exp({-\mathrm{i}\lambda^+_r\cdot x})
		+
		\frac{d^*_r(\vec{\lambda})}{N^-_r(\vec{\lambda})}
		\textrm{\sffamily\itshape w}^{-}_r(\vec{\lambda})
		\exp({-\mathrm{i}\lambda^-_r\cdot x})
		\right]\,.
	\end{equation}
	The complex-valued Fourier coefficients $b_r(\vec{\lambda})$ and $d_r^*(\vec{\lambda})$ 
	may be chosen arbitrarily. 
	This decomposition separates the set of solutions 
	into four terms,
	each of which is defined 
	via coordinate-independent features.

	\section{Quantization}
	\label{quant}
	
	We begin by taking $\psi^1\coloneqq\psi$ and $\psi^2\coloneqq\psi^\dagger$ as Grassmann-valued
	and perform a hamiltonian analysis of our $b_\mu$ theory~\cite{pd50,hrt76,ht94}. 
	The usual equal-time structure is understood throughout.
	The conjugate momenta
	\beq\label{CanMom}
	\Pi^A=\frac{\partial^L\cal{L}}{\partial\dot{\psi}^A},
	\eeq
	where $A\in\{1,2\}$ 
	and $\partial^L$ is the left derivative, 
	permit the construction of the canonical hamiltonian 
	\begin{equation}\label{canH}
		\mathcal{H}_0 
		=\int\! \mathrm{d}^3 x\; (\dot{\psi}^A\Pi^A-{\cal L})
		=\int\! \mathrm{d}^3 x\; \tfrac{1}{2}\mkern+1mu\psi^\dagger\mkern-1mu H\mkern+2mu\psi+\textrm{h.c.}\,
	\end{equation}
	The conjugate momenta determine two primary constraints
	\beq\label{PConstr}
	\chi^A\coloneqq\Pi^{A}+\frac{\mathrm{i}}{2}\sigma^{AB}\psi^B\approx 0\,.
	\eeq
	Paralleling the usual Dirac case, 
	they are second class:
	\beq\label{ConstrComm}
	\{\chi^A(\vec{x}),\chi^B(\vec{x}{\mkern+2mu}')\}=-\mathrm{i}\sigma^{AB}\delta(\vec{x}-\vec{x}{\mkern+2mu}')\,.
	\eeq
	Here, $\sigma=\sigma_1$ is the first Pauli matrix~\cite{iz80}, 
	and $\{\cdot,\cdot\}$ denotes the generalized Poisson bracket.
	
	Conservation of the constraints under time evolution with the extended hamiltonian $\mathcal{H}$ 
	requires
	\beq\label{extH1}
	\{\chi^{A}(\vec{x}),\mathcal{H}\}=\int \!\mathrm{d}^3 x' \; u^B\{\chi^A(\vec{x}),\chi^B(\vec{x}{\mkern+2mu}')\}\,
	\eeq
	for some $u^B$.
	This demands $u^1  =\gamma^0\big(\mathrm{i}\mkern+2mu\vec{\gamma}\cdot\vec{\nabla}+m+\gamma^5\slashed{\phantom{b}}\mkern-10mu b\big)\psi$
	and $u^2=\bar{\psi}\big(-\mathrm{i}\mkern+2mu\vec{\gamma}\cdot\vec{\nabla}+m+\gamma^5\slashed{\phantom{b}}\mkern-10mu b\big)$
	without producing secondary constraints.
	The Dirac bracket therefore becomes
	\begin{equation}\label{DiracBracket}
		\{F(\vec{x}),G(\vec{x}{\mkern+2mu}')\}_D=\{F(\vec{x}),G(\vec{x}{\mkern+2mu}')\}
		-\{F(\vec{x}),\chi^A(\vec{r})\}{\mkern+3mu}(\kappa^{-1})^{AB}(\vec{r},\vec{r}{\mkern+2mu}'){\mkern+3mu} \{\chi^B(\vec{r}{\mkern+2mu}'),G(\vec{x}{\mkern+2mu}')\}\,,
	\end{equation}
	where $\kappa^{AB}(\vec{r},\vec{r}{\mkern+2mu}')=\{\chi^A(\vec{r}),\chi^B(\vec{r}{\mkern+2mu}')\}$.
	Application of Eq.~\rf{DiracBracket} 
	to the fields and their conjugate momenta 
	and explicitly displaying $t$
	yields on the constraint surface
	\beq\label{canDBracketCS}
	\{\psi(t,\vec{x}),\psi^\dagger (t,\vec{x}{\mkern+2mu}')\}_D=-\mathrm{i}\delta(\vec{x}-\vec{x}{\mkern+2mu}')
	\eeq
	with the other Dirac brackets vanishing. 
	
	The above analysis provides support 
	for a quantization procedure 
	that promotes $\psi$ and $\psi^\dagger$ to operators on a Hilbert space 
	and imposes the anticommutators
	\beq\label{CAR}
	\{\psi_j(t,\vec{x}),\psi_k^\dagger(t,\vec{x}\,')\}
	=\delta_{\mkern-2mu jk}\mkern+2mu \delta(\vec{x}-\vec{x}\,')\,,
	\eeq
	with all other equal-time pairings vanishing.
	The corresponding transition to QFT 
	in momentum space 
	promotes $b_r(\vec{\lambda})$ and $d_r^*(\vec{\lambda})$ in Eq.~\rf{planewavedecomp} to operators 
	with complex conjugation replaced by hermitian conjugation. 
	Fourier inversion then yields
	\label{CAR2}
	\begin{align}
		\{b_r(\vec{\lambda}),b^\dagger_{r'}(\vec{\lambda}')\} & = 
		(2\pi)^3N^+_r(\vec{\lambda})\,\delta(\vec{\lambda}-\vec{\lambda}')\,\delta_{rr'}\,, \nonumber \\[1ex]
		\{d_r(\vec{\lambda}),d^\dagger_{r'}(\vec{\lambda}')\} & = 
		(2\pi)^3N^-_r(\vec{\lambda})\,\delta(\vec{\lambda}-\vec{\lambda}')\,\delta_{rr'}\,,
	\end{align}
	with all other anticommutators 
	between these operators vanishing. 
	These relations imply $\{\psi(x),\bar{\psi}(y)\}=0$ for spacelike separations $(x-y)^2<0$, 
	so that microcausality is preserved, 
	a feature compatible with our result~\rf{group_v_bound}.
	
	The algebraic structure of the relations~\rf{CAR2} 
	coincides up to normalization 
	with that of the conventional fermionic creation and annihilation operators. 
	Once the action of $b_r(\vec{\lambda})$, 
	$b^\dagger_r(\vec{\lambda})$, 
	$d_r(\vec{\lambda})$, 
	and $d^\dagger_r(\vec{\lambda})$ on $|0\rangle$ is established, 
	the Fock space may be constructed as usual. 
	To avoid the introduction of preferred frames,
	we require all operators associated to a given branch
	to be of the same type, 
	either creation or annihilation operators.
	To avoid contradictory boundary conditions 
	when interactions with ordinary matter are considered, 
	we select the usual Feynman time evolution.
	
	To elucidate the Fock-space implications of these choices, 
	consider the two-point correlation function $S_{\mkern-4mu F}(x-y)=\langle 0|\mathcal{T}\psi(x)\bar{\psi}(y)|0 \rangle$. 
	Here, 
	$\mathcal{T}$ denotes time ordering, 
	as usual,
	which remains coordinate independent 
	because microcausality still holds. 
	In conventional QFT, 
	the integral representation of $S_{\mkern-4mu F}$
	implements Feynman boundary conditions 
	via an integration contour 
	passing below the negative-frequency pole and above the positive-frequency pole 
	in the complex $\omega$ plane. 
	In the present case, 
	$b_r(\vec{\lambda})|0 \rangle=0$ and $b^\dagger_r(\vec{\lambda})|0 \rangle=0$
	are directly correlated 
	with the contour passing, 
	respectively, 
	above and below the $\omega^+_r(\vec{\lambda})$ pole.
	Similarly, 
	$d_r(\vec{\lambda})|0 \rangle=0$ and $d^\dagger_r(\vec{\lambda})|0 \rangle=0$
	correspond 
	to the contour passing, 
	respectively, 
	below and above the $\omega^-_r(\vec{\lambda})$ pole.
	Thus, 
	the desired Feynman boundary conditions are implemented
	by $b_r(\vec{\lambda})|0 \rangle=d_r(\vec{\lambda})|0 \rangle=0$.
	
	With the above choice, 
	the four number operators are 
	$\mathcal{N}^+_r({\vec{\lambda}})=b^\dagger_r(\vec{\lambda})\mkern+2mu b_{r}(\vec{\lambda})$  
	and $\mathcal{N}^-_r(\vec{\lambda})=d^\dagger_r(\vec{\lambda})\mkern+2mu d_{r}(\vec{\lambda})$.
	The subtracted QFT versions of the conserved 4-momentum 
	and fermion-number operators then become 
	\begin{align}
		\label{2ndMom}
		\!\! P^{\mu}&=\sum_{r}\int \!\frac{\mathrm{d}^3\lambda}{(2\pi)^3}\left[\frac{\mathcal{N}^+_r(\vec{\lambda})}{N^+_r(\vec{\lambda})}(\lambda^+_r)^\mu -\frac{\mathcal{N}^-_r(\vec{\lambda})}{N^-_r(\vec{\lambda})}(\lambda^-_r)^\mu\right]\,, \\
		N&=\sum_{r}\int \!\frac{\mathrm{d}^3\lambda}{(2\pi)^3}\left[\frac{\mathcal{N}^+_r(\vec{\lambda})}{N^+_r(\vec{\lambda})}-\frac{\mathcal{N}^-_r(\vec{\lambda})}{N^-_r(\vec{\lambda})}\right]\,,
	\end{align}
	respectively. 
	Hence, 
	$b^\dagger_r(\vec{\lambda})$ and $b_r(\vec{\lambda})$ create and destroy fermions ($N=1$) 
	with quantum number $r$ and momentum $(p^+_r)^\mu=(\lambda^+_r)^\mu$, 
	whereas $d^\dagger_r(\vec{\lambda})$ and $d_r(\vec{\lambda})$ create and destroy antifermions ($N=-1$) 
	with quantum number $r$ and momentum $(p^-_r)^\mu=-(\lambda^-_r)^\mu$. 
	Note that for antifermions 
	the physical 4-momentum is aligned opposite the associated plane-wave vector,
	as in the conventional case. 
	Thus, 
	all particles have positive energies 
	for sufficiently large 3-momenta.
	However, 
	in any frame with $b^0\neq 0$,
	particles with 3-momenta in a finite-size ball 
	centered at either $+\sqrt{1+m^2/b^2}\,\vec{b}$ or $-\sqrt{1+m^2/b^2}\,\vec{b}$ 
	with negative energies always exist.
	Thus, 
	the occurrence of negative energies cannot in general 
	be remedied with the above quantum-field reinterpretation 
	of the hole states.
	Nonetheless,
	the above analysis demonstrates
	that quantization and Fock-space construction 
	can be achieved
	without the introduction of a preferred frame.

	\section{Identification of the ground state}
	\label{vacuum}
	
	The ground state of a quantum system 
	holds particular physical significance: 
	many observed situations 
	can be treated as perturbations about this state,
	and its stability properties
	are usually taken as key criteria 
	for the viability of any QFT.
	Moreover,
	various physical properties of the ground state,
	such as those arising
	in the contexts of $\theta$ vacua~\cite{Callan:1976je,Jackiw:1976pf} 
	and the Casimir \cite{Casimir:1948dh,Milton:2004ya} 
	and Unruh \cite{Unruh:1976db,Crispino:2007eb} effects,
	continue to be the subject of active research  
	and can have physical implications.
	
	This section discusses concepts 
	relevant for the identification of the ground state
	within the above Fock space 
	of our $b_\mu$ theory.
	To this end,
	we begin by recalling 
	various accepted notions 
	pertaining to ordinary QFTs.
	We then consider these notions 
	in the context of the $b_\mu$ theory,
	finding that physics input beyond that 
	required in the conventional case is necessary.
	We outline the nature of this input 
	and the implications it entails. 
	An analysis of more formal mathematical properties of the ground state
	would also be of interest but lies outside our present scope.
	
	Stability under time translations 
	is often taken as a key feature of a ground state.
	However, 
	strictly within a given Fock space, 
	such as that for our $b_\mu$ theory,
	all states represent free particles 
	and are therefore stable. 
	In this context, 
	stability requirements by themselves 
	are therefore unsuitable for ground-state selection.
	Various additional implicit and explicit assumptions 
	are usually made to proceed.
	
	One is the presence of particle interactions.
	Indeed,
	in many QFT cases of physical relevance,
	Fock spaces merely represent asymptoptic states, 
	and only additional physics
	such as perturbative interactions
	can produce transitions between these states.
	For example, 
	some states may be unstable 
	against scattering processes 
	and can thus be eliminated 
	from the set of candidate ground states. 
	Note also that 
	nonperturbative interactions
	may substantially alter the naive Fock space itself
	and thus the set of candidate ground states, 
	as occurs with quark confinement in QCD~\cite{Shuryak:2004}.
	
	Certain aspects of this stability condition 
	may be captured kinematically 
	because energy--momentum conservation
	can prohibit certain states 
	from transitioning to other states. 
	In the usual Lorentz-symmetric construction,
	all Fock states have momenta 
	\beq\label{forward_cone}
	p^\mu p_\mu \ge 0\,,\quad p^0 \ge 0\,.
	\eeq
	After a suitable subtraction,
	only one state has $p^\mu=0$.
	This state is kinematically stable 
	and thus a candidate ground state.
	However,
	other stable configurations,
	such as any zero-photon single-electron state in QED,
	can also exist.
	To eliminate these states,
	one option might be to require 
	minimal energy for the ground state. 
	However, 
	this choice is {\it ad hoc} 
	and fails to be rooted in physical considerations. 
	Moreover,
	it restricts the zero component of a 4-vector 
	and can therefore only be implemented meaningfully 
	for Fock-space spectra obeying Eq.~\rf{forward_cone},
	a property violated in the $b_\mu$ theory. 
	
	Another type of condition can instead be imposed,
	requiring the ground state
	to remain invariant 
	under the symmetries of the theory.
	This condition is appealing 
	because it is based on physical features of the theory 
	and exhibits more generality. 
	Its discriminatory power
	derives in part from a heuristic observation:
	among all Fock states,
	those with more symmetry 
	occur less frequently.
	In the above QED example,
	single-electron states are then disqualified 
	because they change 
	under various spacetime transformations,
	even though the QED Lagrange density remains symmetric.
	This alternative condition
	retains its relevance in situations involving internal symmetries:
	in a non-abelian gauge theory such as QCD, 
	an infinite set of stable candidate ground states may exist,
	each invariant only under gauge transformations 
	of a given homotopy class~\cite{Callan:1976je,Jackiw:1976pf}.
	Requiring invariance under all gauge transformations
	then selects specific linear combinations 
	within this set. 
	
	We remark 
	that the condition of maximal symmetry
	plays a subsidiary role to stability.
	For example,
	these two conditions are incompatible 
	when spontaneous symmetry breaking occurs.
	In such cases, 
	an infinite set of stable states 
	with reduced symmetry is identified,
	and any member of this set 
	can serve as a ground state.
	The actual ground state is typically fixed
	via a random selection 
	from this set. 
	Spontaneous symmetry breaking 
	also reveals 
	that the intuitive notion of emptiness 
	sometimes associated with QFT ground states
	is too vague to be adopted 
	for ground-state selection. 
	
	In essence, 
	the implementation of the above ideas 
	in the Lorentz-invariant case
	reduces to a symmetry analysis 
	of the theory in question,
	the determination of the subspace of kinematically stable Fock states, 
	and selection of a state from this subspace
	with a maximal set of vanishing quantum numbers. 
	The uniqueness of this state 
	is tied to the presence or absence of spontaneous symmetry breakdown.
	We note in particular 
	that this procedure to identify the ground state 
	is independent of the introduction of external preferred frames
	and also of any specific experimental methodology 
	or physical process 
	that might drive the system to approach the ground state.
	
	To apply the above ideas 
	to our $b_\mu$ theory,
	we begin 
	by allowing general interactions.
	For simplicity, 
	we focus on cases 
	that are not accompanied by additional symmetry violations,
	e.g., 
	by introducing suitable self interactions 
	or coupling to a Poincaré-invariant system 
	conserving $N$.
	Certain single-fermion states 
	can then transition to multifermion states~\cite{kl01}.
	Unlike the conventional Lorentz-invariant case of a single massive species,
	these transitions can occur without interactions 
	involving other degrees of freedom:
	the presence of self-interactions suffices.
	This feature of $b_\mu$ fermions 
	reduces the set of stable single-particle states
	relative to that in a conventional Dirac theory,
	thereby decreasing the number of candidate ground states. 
	Rather than rendering the $b_\mu$ theory inconsistent,
	the instabilities represent a characteristic phenomenological signature 
	for this type of Lorentz violation.
	
	Nonetheless, 
	numerous kinematically stable Fock states can still be identified.
	As one example, 
	fix any frame $\Sigma$ with $b^0\neq 0$,
	consider the negative-energy states 
	identified in the previous section, 
	and fill these states to the horizontal momentum-space plane 
	of constant energy $E=E_-$ with $E_- <0$,
	leaving no other states occupied. 
	When only self interactions and 
	interactions with other degrees of freedom obeying condition~\rf{forward_cone} are allowed,
	this configuration is stable: 
	energy conservation prevents transitions to unoccupied states,
	while Pauli blocking also prevents 
	any fermion transitioning to a lower- or equal-energy state.
	A multitude of stable states of this type 
	can be constructed analogously,
	one for each value of $E_- <0$
	on the dispersion-relation cone.
	
	The energy--momentum plane determining the filling 
	can also be tilted,
	leading to additional stable configurations.
	Such configurations may be found 
	using coordinate independence of the physics.
	The above reasoning in the frame $\Sigma$ 
	also applies in any boosted frame $\Sigma'$:
	filling all states to 
	a plane of constant energy $E'=E'_-$ with $E'_- <0$
	horizontal in $\Sigma'$ 
	also yields stable configurations.
	Note that merely expressing a $\Sigma'$ configuration of this type 
	in the original $\Sigma$ coordinates 
	cannot affect stability properties.
	However, 
	a filling surface horizontal in $\Sigma'$ 
	becomes tilted when expressed in $\Sigma$-frame 
	momentum components $(E,\vec{p})$.
	Kinematically stable multiparticle states 
	in the frame $\Sigma$ 
	can thus include configurations 
	with some empty negative-energy states 
	and some filled positive-energy states.
	This reveals that unfilled negative-energy states
	are not necessarily associated with instabilities.
	
	Next, 
	we search for Fock states 
	that exhibit the symmetries 
	of the Lagrange density, 
	i.e.,
	$N=0$ states invariant under the little group 
	associated with $b^\mu$.
	In a coordinate system 
	with purely spacelike $b^\mu=(0,\vec{b})$,
	it is straightforward to establish 
	that there are five single-particle states with momenta 
	proportional to $b^\mu$:
	$|0\rangle$, 
	a fermion and an antifermion state each with $p^{\mu}$ parallel to $b^\mu$,
	and a fermion and an antifermion state each with $p^{\mu}$ antiparallel to $b^\mu$.
	These states are invariant under the little group
	and can be used to construct six linearly independent states with $N=0$, 
	namely one empty state $|0\rangle$,
	four two-particle fermion--antifermion states,
	and one four-particle state with two fermion--antifermion pairs.
	There are no other $N=0$ Fock states invariant under the little group. 
	Any linear combination of these six states 
	satisfies the ordinary ground-state conditions 
	outlined above.
	
	This theoretical analysis neither leads to a unique ground state 
	nor corresponds to a scenario with spontaneous symmetry breaking.
	However,
	the existence of a definite ground state in Weyl semimetals
	with low-energy excitations governed by a $b_\mu$ theory 
	is experimentally established~\cite{Xu:2015cga,Shekhar:2015,Arnold:2015vvs,Liu:2015tzl}.
	This suggests 
	that the reasoning above is missing a crucial ingredient 
	or needs generalization. 
	
	The key observation is 
	that in the case of a Weyl semimetal
	a general physical procedure is implicitly defined 
	that in principle can be realized experimentally 
	and drives the system to a stable Fock state.
	This state is then taken as the physical ground state.
	The procedure consists of lowering the temperature $T$ 
	of the semimetal, $T\to 0$.
	Note that this physical procedure leans on interactions, 
	as expected from our general discussion above, 
	and in particular on couplings with additional degrees of freedom 
	arising via the dynamics of the semimetal lattice.
	However,
	the procedure also crucially includes thermodynamic considerations 
	and involves the introduction of an external preferred frame.
	
	A detailed discussion of the implementation of this idea 
	in the context of our $b_\mu$ theory 
	can be found elsewhere~\cite{WeylConcordance}. 
	We restrict ourselves here 
	to a few conceptual implications.
	To this end,
	we continue to assume $N$-conserving interactions 
	with additional Poincaré-invariant degrees of freedom. 
	Treating the system at finite temperature 
	then generally introduces additional symmetry violations 
	in the form of a rest frame,
	which represents a preferred frame
	to which we may assign a timelike 4-vector $X^\mu=(X^0,\vec{0})$.
	It is then natural to expect the ground state
	to depend on both $b^\mu$ and $X^\mu$. 
	Different relative orientations between $b^\mu$ and $X^\mu$,
	i.e., different external preferred frames,
	may then entail different ground states.
	
	Situations in which the ground state of a dynamical system 
	depends on externally prescribed quantities 
	are common in physics. 
	Consider, 
	for example, 
	the production of a cylindrical electret~\cite{Gutmann:1948} with axis $\hat{z}$.
	Depending on the orientation of the external electric field $\vec{E}$ 
	relative to $\hat{z}$ during cooling,
	electrets differing in polarization relative to $\hat{z}$, 
	i.e., different ground states, 
	may be achieved.
	In a Weyl semimetal,
	however,
	this effect is obscured
	because $b^\mu$ and $X^\mu$ are inextricably linked via the lattice.
	Their relative orientation, 
	and thus the ground state, 
	are therefore essentially frozen.
	Similarly,
	viewing the $b_\mu$ theory as fundamental,
	the ground state must also depend on external physics 
	involving a preferred frame specified by $X^\mu$. 
	Purely theoretical input 
	is insufficient to select that preferred frame 
	and hence to fix a unique ground state.
	For prospective applications in fundamental physics,
	however,
	$X^\mu$ is already established in the context of cosmology
	as the rest frame of the Big Bang,
	via interactions with the thermal bath
	of Standard-Model and gravitational fields.
	For phenomenological purposes,
	the identification of the ground state can therefore be unambiguous. 
	
	Our analysis of the $b_\mu$ theory has shown 
	that quantization and Fock-space construction 
	can be achieved without reference to an external preferred frame,
	and that ground-state identification 
	based on conventional wisdom without such a frame fails.
	Existing physical systems suggest 
	a resolution of this issue 
	involving preferred-frame input 
	in the form of thermodynamical considerations. 
	This insight provides the basis 
	for more detailed explorations 
	of the longstanding concordance problem~\cite{WeylConcordance}
	and offers intriguing prospects for future discovery.

	\section*{Acknowledgements}
	
	This work is supported in part
	by the U.S.\ Department of Energy
	under grant {DE}-SC0010120,
	by grants FAPEMA Universal 00830/19
	and CNPq Produtividade 310076/2021-8,
	by CAPES/Finance Code 001,
	and by the Indiana University Center for Spacetime Symmetries, 
	College of Arts and Sciences, 
	and Institute for Advanced Study.

\end{document}